\documentclass{IETpaper}

\usepackage{microtype}

\begin{document}


\title{Respectful Things: Adding Social Intelligence to `Smart' Devices}

\author{Max Van Kleek, William Seymour, Reuben Binns, Nigel Shadbolt}

\institution{Department of Computer Science, University of Oxford, Oxford, UK, OX1 3QD\\
\{max.van.kleek, william.seymour, reuben.binns, nigel.shadbolt\} @ cs.ox.ac.uk
}

\maketitle

\begin{keywords}
Respect, smart homes, IoT, privacy, profiling, ethics 
\end{keywords}

\begin{abstract}
In this paper, we propose that the idea of devices respecting their end-users may serve as a strong design goal for highly personal and intimate smart devices. We ask what respect is, how it shapes interaction, and how good-faith simulation of respect might inform user-friendly smart device design. Respect is a natural and integral part of natural human relationships that is seen to shape work and personal relations. In a basic sense, this is the core purpose of smart things: we expect them to be ready and willing to help us. In this vein, we distill the characteristics of more complex respectful behaviours into 4 main types relevant to smart devices, drawing from philosophical analyses of the conceptual dimensions of respect: directive respect, obstacle respect, recognition respect, and care respect. We discuss the implications of each of these kinds of respect for the future of smart personal devices.
\end{abstract}

\section{Introduction}
Inspired by Asimov's Three Laws of Robots, AI ethicists have contemplated whether there should be a set of fundamental principles should guide the design and development of various kinds of robots. The EPSRC, for example, published ``5 Principles of Robotics'' as the output of a workshop involving robotics researchers and social scientists, based roughly around safety, lawfulness, honesty, and responsibility ~\cite{boden2011principles}.
An equivalent discussion around the "ethics of smart IoT" is merited, especially as intimate devices become entrusted to very private spaces, such as in one's home (including one's bathroom, and bedroom), on one's body (e.g. wearables), or even in one's body (e.g. implantables).  Viewed as simple robots themselves, capable of reasoning, learning, and autonomous action, principles for ethical robotics, such as those previously mentioned, could be seen as relevant to such devices.  Yet, there is something about the intimacy of these devices, and the roles they play in people's lives, as well as from  the very privileged access they gain to users' lives, that could be seen as demanding altogether different kinds of design guidelines, to enable positive and responsible relationships with people. 
 In this article, we propose that that this might be be embodied in a single, simple concept: respect, and that the idea of devices respecting their end-users may serve as not only a viable, but a strong goal in the design of highly personal or intimate smart devices. We start by asking what respect is, and how it shapes the ways people interact and behave, before tackling the issue of how respect might translate to the design of smart devices.

\section{Respect in Human Relationships}

Respect is a natural and integral part of human relationships, both professional and personal. Respect for others is also an essential skill in social cognition, alongside courtesy, tolerance of others, and other pro-social attitudes and values.  Perhaps most importantly, being respected is often intricately related to other critical ways people relate to one another, such as trustworthiness, interest, care, and affection.

Thinking about how respect is implicated in human relationships may illuminate ways to reconsider common principles appealed to in the context of Internet-of-Things. Take, for instance, trust. In such contexts, trust is often used in the sense of `trusted computing', where secret cryptographic keys are used to enforce that certain processes operate as intended by the manufacturer. However, in human relationships, trust is not \emph{enforced} but rather \emph{built up} as part of a relationship between two or more entities. Trust is only an appropriate stance if the trusted thing is in fact \emph{trustworthy} ~\cite{o2014trust}.

Trust as a social phenomenon is also distinguishable from mere \emph{reliance} ~\cite{baier1986trust}, in ways that relate to respect. If one merely \emph{relies} on the actions of somebody or something (such as an alarm clock bleeping at the right time, or someone else pressing the right button in an elevator), one may be disappointed if that person or thing fails to behave as expected, but one wouldn't feel betrayed, nor would one lose respect for them / it. But if one ~\emph{trusts} someone, and that trust is violated, one may rightfully feel betrayed, and subsequently lose respect for them. Such emotional stances would not make sense if directed towards one's alarm clock, or towards a stranger in an elevator. Violations of trust diminish respect, and respectful behaviour begets perceptions of trust-worthiness. Thus, respect and trust are inter-related concepts which only make sense in a network of social-emotional interactions. In these senses, trust and respect are not measurable features of a technical system, but rather emergent properties of relationships.

Respect can also have other effects beyond building mutual trust. For example, good bedside manner by clinicians, which includes being respectful, has been associated not only with better patient satisfaction, but with measurably better outcomes.  Part of this has been explained in terms of respect fostering a better shared understanding of a person's needs and preferences---an understanding that is a prerequsite for such needs and preferences being met. Thus, bedside manner is seen as important to achieving quality care in healthcare delivery~\cite{weissmann2006role}.  

Healthcare is also a setting which exhibits useful examples pertaining to the need to make delicate tradeoffs in  respectful interactions. For example, while mutual openness and honesty may be seen as important characteristics of respectful relationships, it is well-established that doctors will not be fully open or truthful to patients when doing so might cause them harm. Such cases include telling patients their treatment success odds, true estimates of expected life expectancy in palliative care, or disclosing the extent of the decline in a patient's condition, including cognitive decline~\cite{matthias2015robot}.  Such examples illustrate that respect often requires carefully considered choices and the implications that such choices have on the other's well-being.




\section{Types of Respect}







What does it \emph{mean} to have respect, and does it comprise one or a set of specific considerations, views or behaviours?  Many writings in analytical philosophy have attempted to answer such questions and to unpack our intuitive understandings of respect. In this paper, we focus on four characterisations most relevant to smart devices: \emph{direct respect}, \emph{obstacle respect}, \emph{recognition respect}, and \emph{care respect}. When talking about respect in this article we use the notion of the subject of the respect, who is the entity doing the respecting, and the object of the respect, which is the person or item being respected. Rather confusingly then, here we are talking about smart devices being the subject of respect, and users as the object of respect.

Direct respect encompasses rules and requests which specify a desired behaviour. For a subject to show direct respect to an object, then, means for the subject to voluntarily comply with the object's rules in the spirit that they are intended. This is the most basic type of respect, and is often the first that children come to understand. The only potential room for maneuver is in the interpretation of requests that have a discretionary component, otherwise the choice is simply to obey the command or not. In the context of a smart home IoT device, the device (the \emph{subject}) might be said to respect the configuration preferences of the user (the \emph{object}).

Obstacle respect, on the other hand, can be thought of as the subject regarding the object in a particular way on the basis that not doing so might hinder the subject’s ability to pursue its own ends. This captures the precarious trade-off that is present in devices that treat the customer as a product, either through advertising or data mining. Acknowledging that part of the respect that such devices show is as a means to generate profit should be an integral part of the process of accepting such a device into the home, especially when profit making runs antithetical to other goals the device may have (such as respecting the privacy concerns of the user); exchanges involving information and other intangible rights should be made explicit before they are accepted.

But none of these definitions so far capture the notion of a device altering, unbidden, how it behaves in relation to the user. Instead of having to formally define the boundaries that exist between device owners and other people a smart device may be able to infer this after observing interactions between them over time. We can use the term recognition respect to describe this, and it represents a step away from the idea of a mindless machine towards what one would expect from a respectful human.

Finally, we come to the type of respect that involves treating someone in a way that genuinely supports their wellbeing rather than merely insofar as it serves one’s own self. Care respect describes a relationship where the respect giver takes on a caring role as a result of seeing the object as having intrinsic value. They might prioritise the long term goals of the object over short term objectives in an effort to do `what’s best for it’, with no consideration for their own concerns. The need to care for a user has to be carefully balanced with respect for their autonomy, and should ultimately be subservient to their ability to understand and influence their own circumstances.

Of course, respectful relationships are composed of many different types of respect in varying quantities. This complexity is what makes human relationships fulfilling and will be important in order to successfully integrate socially aware smart devices into the home. In order to accomplish this we need to have an understanding of both how device makers might feel towards their users, as well as what devices need to know and do in order to exhibit these flavours of respect.

\section{What Would It Mean for Smart Devices to Be `Respectful'?}

The idea of respecting users as a desired characteristic leads first to the question of what having respect does to the relationship of the device and an individual. According to the Stanford Encyclopedia of Philosophy, ``when we respect something, we heed its call, accord it its due, acknowledge its claim to our attention'' ~\cite{sep-respect}. In a basic sense, this is the core purpose of smart things: we expect them to be ready and willing to help us (and would be far less inclined to use them if they were antagonistic). So at a certain level one could say that all such devices were already designed with some amount of respect in mind.

What is the opposite of being respectful? Acting selfishly towards someone, being manipulative or, an extreme sense, Machiavellian. But beyond this simple example there are many types of respectful behaviour. For instance, there not only exists a relationship between a smart device and its owner, but also between the device and the other people who share the same social space. To be respectful in this context would require a device to have a level of social awareness and ability to respect personal boundaries. Its interactions with third parties would be mediated by the relationship that exists between the third party and the owner of the device (e.g. it is probably inappropriate to enumerate the contents of the owner's calendar to a visitor, for example). We might imagine a respectful device to initially be more withdrawn, slowly ‘warming up’ over time as it learns who the user trusts with different kinds of information.

There are several reasons why respect has not previously been viewed as a kind of short-term design goal for systems; one primary difficulty is that there has historically been a view that no non-human entity could ever truly have respect given that, for some definitions, respect transcends action or behaviour, and relates to how a person sees, views, and feels about another person or entity. The conclusion being that although animals (and smart devices?) may love or fear us, only people can respect and disrespect us or anything else ~\cite{sep-respect}.

Yet putting aside the argument about cognitive states, or what it means to have feelings, there is little in this argument that would preclude smart devices from being able to exhibit behaviour that suggests such feelings or states. Here we draw the distinction between the ability to have respect for something, and the ability to perform actions which show, or are normally symptomatic of, respect.

So while (epistemologically speaking) a device may only be capable of showing respect to an end user through the instructions given to it by another human, there is no functional difference in its behaviour compared to something which is truly capable of having respect. Indeed, this difference does not prevent users from anthropomorphising smart devices, treating and thinking about them as human or possessing human characteristics. And when we consider the inverse, the ability to show disrespect, smart devices are certainly capable of being obtuse, annoying, and intrusive, and they are held responsible too; if a device fails to respect the user then it will be removed from the house and disposed of.

 As such, whether we conceptualise the respect as being had or shown, it remains an important tool which designers can use to improve the usefulness of the devices they create. But this does prompt the question as to what it means for a smart device to be self-absorbed, or have egocentric concerns and motives. Device manufacturers might instill certain tasks, goals, drives and priorities into the device that cause it to serve its own needs. In order to know that the device was respecting them, how sure would users need to be that their devices were prioritising their own needs over that of their creators? This is possibly the greatest barrier preventing users from trusting smart devices.


\section{Designing for Respectful Interaction}

What are some potential implications of designing emerging smart devices and tools to be respectful for designers?  In this section, we discuss the kinds of ways we might interpret respectfulness in the context of the kinds of respect just discussed.

Starting with direct respect, we might say that that evidence of respect in  design of smart devices that provide settings such as permissions that constrain their behaviour in particular ways. Even at this basic level, devices can fail to respect users' preferences, such as by failing to make certain potentially objectionable behaviours optional or curtailable, or by obscuring such behaviours so that they are invisible and inscrutable.  

We can begin by tackling direct respect in the context of data protection law and regulation. There already exist regulations such as the GDPR which oblige device producers to respect certain rights that users have with respect to their personal data. Obviously, following rules such as these forms the baseline of what we can expect from smart things, and will indeed be the minimum required in order to make them legally saleable. This also applies to preferences that users might have, particularly with respect to data sharing and important sociocultural boundaries which devices must respect in order to be permitted into the home.

The next logical step is to have devices move beyond honouring explicitly expressed preferences towards embodying more general values that the user holds. This is particularly important as devices become increasingly personal, in some cases acting like an extension of the user. For example, a device could automatically curtail or expand analytics reporting according to how much the user was perceived to care about privacy making informed choices, acting according to observed norms instead of declared ones. 

But what if Apple required that Siri send back data, even though the user was completely set against any sharing of their personal information? If devices prioritise the concerns of their users over those of their creators, we can imagine a situation where they go “rogue” and returned mangled or dummy data, or perhaps even pretend to be switched off. This takes the possibility of being disrespectful described above and extends it to the more complex types of respect described above, which may more accurately represent the type of smart device that a user might want to share their home with.

The notion of respectful devices covers more than just an individual's privacy with respect to data sent to cloud services the device communicates with. We can expect socially adaptive ‘things’ to be considerate of personal boundaries inside the home; instead of a user having to explicitly state to a device what information to share with each person who enters the house, it would be much more respectful to observe the types of interactions which take place between third parties and the owner and emulate this with new arrivals. In order to prevent potential disrespect whilst the device learns about the user, it would be preferable to have devices start closed/restricted and share more over time, with any tradeoffs being made explicit (Siri might inform a user who has requested a translation that unlike normal speech recognition, this involves sending their raw speech waveforms to Apple).

This idea of socially adaptive software can be taken further, by having smart devices mediate the flow of information from and to the end-user. For example, by automatically paraphrasing or providing content warnings, a device might spare its user unnecessary aggravation from messages or posts that target their specific sensibilities. Similarly, a device might choose to lie in order to help its owner save face during a difficult social situation. This emulation of how people not only constantly attempt to repair their own images in the view of others, but also engage in such reparative activities even for others, especially those of friends, colleagues and allies would build on the bond between user and device. Given the idea of respect, thus, one might argue that people and things may be naturally inclined to proactively help to save face and make reparations.

Finally, when these decisions are taking place on the device itself, and are the result of learning in addition to the device’s original programming, one must ask the question of who would be responsible for decisions made by a smart thing (particularly disrespectful decisions). The intuitive argument that the creators should be held responsible begins to break down as the device learns and adapts to its owner, with the responsibility held by the creator slowly diminishing over time in the same way that a parent eventually ceases to be responsible for their child.

This line of reasoning might lead us to hold the owner of a device partially responsible, since it is them that the device is learning from. But establishing any kind of causal link between actions performed by the user and subsequent actions performed by the device is near impossible for all but the simplest of devices. Distributing responsibility across the creator and the user is the worst possible outcome, with so many people being involved that everybody’s problem becomes nobody’s responsibility.


But unless we are willing to ascribe responsibility to the smart device itself, this leaves a responsibility gap, where nobody is assigned responsibility for the actions of a smart device. While this may be acceptable when a smart assistant suggests an outfit where the colours clash, one can easily imagine situations where devices failed to call for help for elderly users, or decided against informing a user’s friends of a potentially abusive relationship.

\section{Discussion}

In this section, we first briefly touch on a variety of specific challenges to achieving
truly respectful systems, framing them primarily as questions and areas of potential
further investigation.

\subsection{Loyalty \& the Complexities of Multi-Party Respect}
With any widespread adoption of technology which mediates relationships between people comes the risk that it will systematise and normalise unfair power dynamics. Consider the scenario where a family in a shared house\footnote{\textit{House of Multiple Occupancy} in the UK} purchases a smart device. How should the loyalties of the device be split? In such a situation there are multiple relationships and power dynamics which could be exploited through a device which obeyed the letter, if not the spirit, or respect.

Within the family group, how should the device behave towards each member of the family? Should partners be afforded the same permissions and use of the device as each other, or should the device allow them to keep secrets? Either choice could lead to tensions between them depending on familial norms. One could argue that such problems would exist without the presence of the device, but this would ignore the role of the device in normalising the behaviour (in the same way as smartphones have normalised the behaviour of not paying attention to others whilst in their company). 

A device could also impact the relationships between tenants relationships in the house. While at first it might seem that it would be disrespectful for others outside the family to have use of the device, they have an equal right to use the common spaces in the house. The refusal of a device to heed their commands (to stop playing loud music, for example) inhibits that right. It was mentioned above how being respectful to the user might entail ignoring the goals of device manufacturers, but arguing that other \textit{people} sharing the same space should be so ignored.

Finally, what about the relationship between the family and the landlord of the house? Should a respectful device owned by the landlord be able to collect information on other people living in the house? Clearly being physically co-located with a device grants a basic level of respect which supersedes that of the owner, but it might be difficult to persuade consumers to accept a loss of agency, when the purpose of the respectful smart device is to be empowering.




\subsection{Respecting Devices}
It is important to distinguish devices having respect for users with users having respect for devices. In prior work it has been observed that, like when interacting with other people, users tend to act more politely to machines which ask the user to rate the machine in question than when they are asked about the machine by a person (or even on another identical machine) ~\cite{nass1999people}. Studies such as this are part of a wider body of work in in which some argue that robots (and by extension smart devices) should be considered sentent, moral agents ~\cite{bryson2010robots}, others that their status should be similar to that of slaves ~\cite{hern2017give}, and everything in between.

The purpose of this work is not to answer these questions, nor to consider how users should treat their devices (and whether them being respectful should have any bearing on this). Put another way, instead of wondering if people should treat machies like people, this paper only assume that machines should treat people like people (a much easier argument to make).




\subsection{Devices Gaining a User's Trust}
What if being passively respectful (as has been advocated up until now) is not enough? Generations of devices and services which have been disrespectful, selling or leaking user data might have led to users being suspicious of smart devices which claim to be respectful.

In such a situation, what might a device have to do in order to prove itself to its owner? An obvious answer might be to curtail the reporting of personal or usage data to the manufacturer, but this is not something that can be verified by the user. If it would not adversely impact functionality, it could be possible for a smart device to suggest that it be temporarily disconnected from the internet (but not the home network) to demonstrate that it prioritises its user over its creator.

This behaviour could be triggered by observations of the user. Perhaps, after several weeks of use, the user has not given much information to the device, or has not connected it to other devices in the house. This could prompt the device to suggest that it prove its loyalty as a way of advancing the user's long term interests (by giving them a better experience in the future).




\subsection{The Political Economy of Respectful Things}

Whether there will be devices made that are truly respectful depends hugely on the many kinds of unseen factors that influence the design of products and systems today.  For commercial products and services, for example, pressures such as the desire to drive continued user engagment, or getting ``conversions'' on placed advertisement often influence the design of interface and interaction at multiple levels.  The incentives that drive such pressures have been so great that they have spurred the growth of an industry of  ``dark patterns'' of interaction~\cite{zagal2013dark,greenberg2014dark} that   apply behavioural economics principles to shape and nudge users' behaviour without their explcit consent or even awareness, so as to minimise the likelihood of objection or that users will actively contravene.   Such patterns have been seen as responsible for the rise of app addiction, fake apps and deceptive games that trick people into in-app purchases~\cite{soroush2014self}, as well as other phenomena such as growing online conflict and political polarisation~\cite{marichal2016facebook}.

Such patterns should be perceived as disrespectful, using the proposed framework for respect, for at least two reasons; not only are they a form of deception in which users are being coerced to behave in particular ways without their explicit knowledge or consent, but many of these behaviours may be directly at odds of individuals' personal goals, as well as those of the societies and groups to which he or she belongs.  Such personal goals might be, for example, to not spend too much time on social media, to avoid certain kinds of content, or to avoid buying things they don't need, while group goals might be to remain politically neutral, tolerant of others' views, or to promote civil discourse and progress.

Beyond the interface, there may be pressures that have shaped the fundamental architectures of the platforms upon which apps, services, and IoT devices are being built that are also odds with people's desires and needs.  One has to do with unwanted or unexpected data collection; in particular, the pervasive model of behavioural ad targeting has driven a thirst for user data, leading apps and service developers in a race to fuel the so-called ``surveillance economy'' of user data exchanged on ad networks~\cite{zuckerman2014internet}.  Such a race has, in turn, caused platforms and service developers to make such choices as adding data collection where it is not needed for functionality, to disguise data collection efforts where possible, and to tie as much functionality to central cloud services controlled by these platforms so as to cause as much data to be sent to them as possible.  Even less visible are the ways in which such services exchange and repurpose potentially sensitive user data through brokers and exchanges, as well as through back-room relationships with ``partners''.

Beyond data harvesting and repurposing, these architectures have made design decisions that can be seen as disrepectful by limiting user choice, freedom, or autonomy.  One example is inconveniencing users by limiting interoperability with competitors' services.  Examples abound, but one particularly simple example is how Facebook Messenger and Google Talk  discontinued support for external integration with standard protocols (such as XMPP) that enabled its users to communicate with those users of competitors' services.  Another example are high barriers to exporting data from platforms that might facilitate and thereby encourage users to leave them.

Some may argue that the only way to that such services and platforms could be made independent of such profit-driven design choices is for them to be developed by non-profit organisations and foundations.  Indeed, much of the core infrastructure of the Internet, along with commercial platform infrastructure is powered by truly vendor-agnostic free and open source (FOSS) software projects which are themselves contributed to by many for-profit organisations.  The global visibility of contributions to such tools, however, ensures that design choices within them are not made to benefit one party over other, respecting all users equally.  In order for networked end-user applications and services to respectful, there may be the need to build  virtual infrastructure using such neutral FOSS tools, to replace those services that have disrespectful design assumptions built-in.  

Still other design obstacles might be presented by legal obligations or interactions with national governments.  For example, laws restricting the distribution of certain kinds of material might force even respectful things of the future to have to violate users' instructions, or, to turn over private data against their owners' wishes based on requests from law enforcement.  Data captured by Amazon's Alexa has already been used as evidence in a murder trial~\cite{alexamurder}; if such data had been collected by a respectful device, would the device have destroyed the data---simply because, respectful of the user's privacy, it was only keeping data as needed, or because it identified its potential sensitivity---before it could be found by law enforcement?  If required by law, to keep such data, would Alexa-like devices in the future need to identify potential crimes in data collected to identity where data needed to be kept?

\section{Conclusion}

Once we begin to move beyond the capabilities of current ‘smart’ home devices towards those which are more socially aware and occupy more intimate places in our homes, we need a conceptual understanding of the properties we want these devices to have in order to shape norms and positively influence how the market develops. While the notion of respect broadly covers what we are looking for, we have shown in this article that respect has a number of different meanings that are appropriate to different situations in different ways.

Respectful devices which are not trusted are of little use, and enabling users to verify that their devices are being honest is potentially the largest problem facing the introduction of smarter devices in the home. As long as the architectures which power smart-devices remain centralised, solving this seems like an impossible task. Decentralised data storage, on the other hand, might offer a way for users to keep devices honest by examining the types of data which leave their local networks and laying the building blocks of trust.

Leaving aside philosophical discussions about the moral capabilities of smart devices, it is evident that even if we do not truly believe that smart devices are capable of being respectful or responsible, we do need to account for the way that they will be perceived by end-users. Indeed, certain types of respect need to be shown by devices, like humans, in order for them to successfully co-exist in a home environment. Beyond this, showing respect is not only a social good, but will also engender more trust in users, which may in turn either directly or indirectly further the goals of device creators.

\section*{Acknowledgements}
This work was supported by \emph{ReTiPS: Repectful Things in Private Spaces}, a project funded through the PETRAS IoT Hub Strategic Fund, which, in turn, was funded by the UK Engineering and Physical Sciences Research Council (EPSRC) under grant number N02334X/1.



\bibliographystyle{plain}
\bibliography{main}

\end{document}